\documentclass[reprint,twocolumn, prl]{revtex4-1} 
\usepackage{graphicx} \usepackage{epsfig} \usepackage{color} \usepackage{braket}
\usepackage{amsmath,amsfonts,amsthm,bm}
\usepackage[normalem]{ulem}
\newcommand\redout{\bgroup\markoverwith
{\textcolor{red}{\rule[0.5ex]{2pt}{0.8pt}}}\ULon}
\usepackage[ breaklinks,colorlinks,
linkcolor=blue,citecolor=blue,urlcolor=blue]{hyperref}
\begin{document}
\title{
Yang monopoles and emergent three-dimensional topological defects
in interacting bosons
}
\author{Yangqian Yan,  Qi Zhou}
\affiliation{Department of Physics and
Astronomy, Purdue University, West Lafayette, IN, 47906}
\date{\today}
\begin{abstract}
   Yang monopole as a zero-dimensional topological defect has been well
established in multiple fields in physics.  However, it remains an intriguing
question to understand interaction effects on Yang monopoles.  Here, we show
that collective motions of many interacting bosons give rise to exotic
topological defects that are distinct from Yang monopoles seen by a single
particle.  Whereas interactions may distribute Yang monopoles in the parameter
space or glue them to a single giant one of multiple charges,
three-dimensional topological defects also arise from continuous manifolds of
degenerate many-body eigenstates.  
Their projections in lower dimensions lead to knotted nodal
lines and nodal rings. Our results suggest
that ultracold bosonic atoms can be used to create emergent topological
defects and directly measure topological invariants that are not easy to access
in solids. 
\end{abstract}

\maketitle
Yang monopoles play a crucial role in non-abelian gauge theories
and have influential impacts in multiple subareas of physics~\cite{yang54}. 
In high energy physics,
they lay the foundation of Yang-Mills theory and standard
model~\cite{yang78,thooft07,thooft74,chern1,chern2}. In condensed
matter physics, they give rise to nontrivial topological quantum
states characterized by the second Chern number,
$C_2$~\cite{zhang04,chern0,qi08}.  In a 5D parameter
space, a Yang monopole represents a zero-dimensional point defect with a
four-fold degeneracy. Away from a Yang monopole, a spin-3/2 or pseudospin-3/2
could see such point topological defect from either local non-abelian Berry
curvatures or $C_2$. When a 4D surface encloses the Yang monopole, $C_2=1$.
One could view a Yang monopole as a magnetic monopole of ``charge" 1.

Whereas Yang monopoles remained a theoretical concept for decades, Sugawa, {\it
et al.}, at NIST delivered a Yang monopole for the first time in laboratories by
engineering the couplings among four hyperfine spin states of ultracold bosonic
atoms~\cite{nistarXiv}. Each boson in this experiment represents a pseudospin-3/2.
While many experiments have used bosons to probe local abelian Berry 
curvatures~\cite{zakexp13,butterfly15,bloch16}, $C_2$ has been extracted 
in the NIST experiment by integrating the non-abelian Berry curvature on 
4D surfaces.
Very recently, $C_2$ has also been measured in optical
lattices and photonic crystals~\cite{bloch17,rechtsman17}.

Though Yang monopoles have been well established in non-interacting systems,
a fundamental question remains. 
Are topological defects seen by a collection of many interacting spin-3/2s 
the same as those seen by each individual one?  In
this Letter, we show that interactions allow physicists to access completely
different 
topological defects arising from collective motions of many particles. These
emergent topological defects signify the vital importance of interactions on
Yang monopoles, and demonstrate the power of ultracold 
atoms in creating
and detecting novel topological phenomena that are not easy to access in solids. 

Our main results are summarized as follows.
For odd particle numbers $N$, repulsive
interactions distribute Yang monopoles 
on a quantization axis
in the parameter space, and attractive interactions glue them 
to a single one of ``charge"  $N^2$ at the origin.
In contrast, for $N=4n+2$, where $n$ is a 
non-negative integer, interactions
produce multiple 3D topological defects. 
When $N=4n$, the many-body ground state is unique for repulsive interactions,
and no topological defect can be seen by the ground state.  The results of
attractive interactions are similar to those for $N=4n+2$. Here, 3D defects
emerge purely from interaction effects in bosons, unlike those
studied in non-interacting electronic
systems~\cite{nodal0,nodal1,nodal2,liang15,lian16}.
We also show how to use ultracold bosons to directly measure the topological
invariants in laboratories. 

Our work was motivated by a recent paper by Ho and Li~\cite{liarXiv}. Based on a
mean field approach, this pioneering work shows that a Yang monopole 
may be stretched into an extended manifold due to interactions. 
In this mean field approach, all pseudospin-3/2s
are described by the same condensate wave function.  Here, we
provide an exact solution for a generic $N$ pseudospin-3/2 system.
We show that the many-body
ground state becomes degenerate in certain locations in the parameter space.
These degenerate many-body eigenstates give rise to 
novel topological defects beyond mean field predictions.

{\it Hamiltonian.} The single-particle
Hamiltonian that describes a Yang monopole reads~\cite{lian16,liarXiv}
\begin{equation}
  \hat{K}=-R_z \tau_z\otimes  \hat{n}\cdot \vec{\sigma} -R_x\tau_x-R_y \tau_y,
  \label{spinmodel} 
\end{equation}
where $\vec{\sigma}$ and $\vec{\tau}$ are two
spin-1/2 operators, and $\hat{n}$ is a unit vector.
A single-mode approximation has been taken for the orbital part of
the wavefunction, i.e., bosons share the same spatial wavefunction. 
Eq.(\ref{spinmodel}) defined
in a 5D parameter space, ${\bf R}=(R_x, R_y, R_zn_x, R_zn_y, R_z n_z)$,
describes a spin-3/2 particle. 
In the NIST experiment, the first (second) two parameters are determined by the
intensities and phases of radio-frequency (micro-wave) coupling and the last one,
$R_zn_z$, is adjusted 
by the detuning~\cite{nistarXiv}. 
For convenience of later discussions,
We rewrite this Hamiltonian,
\begin{equation}
  \hat{K}=\sum_{i=1}^4
  \epsilon_i \hat{a}_i^\dagger\hat{a}_i+\sum_{j=2}^4\sum_{i=1}^{j-1}(t_{ij}
  \hat{a}^\dagger_i\hat{a}_j+h.c.),
  \label{4site} 
\end{equation}
which
describes four lattice sites coupled by certain inter-site tunnelings
$t_{ij}$, as shown in Fig.~\ref{fig1}(a).  
\begin{figure} \centering
  \includegraphics[angle=0,width=0.48\textwidth]{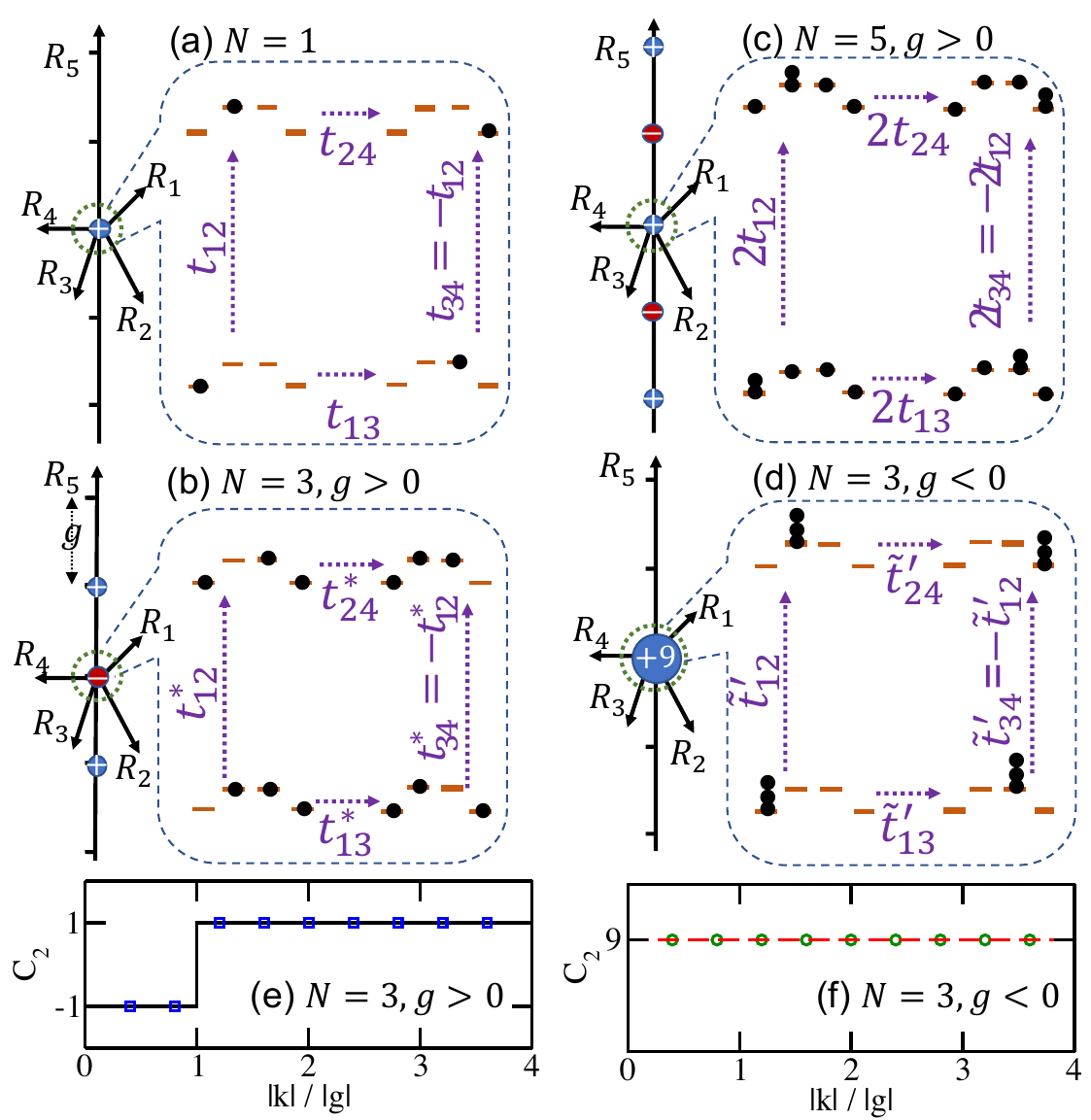}
  \caption{ 
    Yang monopoles for odd $N$.
    Blue (red) spheres show the positively (negatively) charged monopoles
    with charges denoted. 
    (a-c), $g>0$, $N=1, 3, 5$. (d) $g<0$, $N=3$. 
    Insets illustrate effective Hamiltonians near the origin.
    Orange solid lines (black dots) represent single particle states (bosons). 
    Dotted arrows show effective couplings.
    (e-f) $C_2$ as a function of the radius $|k|$ of the 5D sphere
    for three particles.
    Solid and dashed lines (squares and circles) are analytical
    (numerical) results.  
  }
  \label{fig1}
\end{figure}
$\hat{a}_i^\dagger$ ($\hat{a}_i$) is the creation (annihilation) operator at
site $i$.
$\epsilon_i$ is the onsite energy,
$-\epsilon_1=\epsilon_2=\epsilon_3=-\epsilon_4=R_zn_z$,
$t_{13}=t_{24}=-R_x+iR_y$,
$t_{12}=-t_{34}=-R_zn_x+iR_zn_y$, and $t_{14}=t_{23}=0$.
The 5D parameter space is now spanned by the complex tunnelings $t_{12}$ and
$t_{13}$, and $R_zn_z$ that determines the onsite energies.
As this pseudospin-3/2 Hamiltonian respects time reversal symmetry,
every eigenstate is  doubly degenerate, consistent with Kramers theorem. 
For many-body systems, we consider the Hamiltonian,
\begin{equation}
  \hat{H}=\hat{K}+\hat{U}, \,\,\,\,\,\, \hat{U}=g
  \sum_{i=1}^4(\hat{a}_i^\dagger a_i)^2
  =g\sum_{i=1}^4n_i^2
  \label{int}
\end{equation}
where
$g$ is the onsite interaction strength. 
$n_i$ represents the occupation in the $i$th lattice site and
$\sum_{i=1}^4n_i=N$ is satisfied. 
Because $\hat{U}$ respects the time reversal symmetry,
for odd $N$, the Kramers theorem still applies.
Though inter-spin interactions exist in the NIST experiment,
here, we concretize the discussions on intra-spin interactions,
which corresponds to onsite interactions in Eq.(\ref{int}),
to reveal fundamental interaction effects on Yang monopoles.
A Yang monopole may
also be realized alternatively using coupled four lattice sites
described by Eq.(\ref{4site})~\cite{supnote}.
It is then natural to consider $\hat{H}$
in Eq.~(\ref{int}) as onsite interactions dominate. 

We solve $\hat{H}$ exactly and obtain the many-body eigenstates
$|\Psi_m\rangle$ for $N$ bosons. $|\Psi_m\rangle$ is expanded using Fock
states,
$|\Psi_m\rangle=\sum_{\{n_i\}}{\alpha}_{m\{n_i\}}|n_1,n_2,n_3,n_4\rangle$.
When nodal points are
observed, we compute $C_2$~\cite{chern1,chern2,supnote},
\begin{equation}
  C_2=\frac{1}{32\pi^2}\int_{S^4} d {\bf
  R}\epsilon_{\mu\nu\rho\lambda}(\text{Tr}
  [F_{\mu\nu}F_{\rho\lambda}]-\text{Tr}[F_{\mu\nu}]\text{Tr}[F_{\rho\lambda}]),
  \label{C2}
\end{equation} 
where $F_{\mu\nu}=\partial_\mu
A_\nu-\partial_\nu A_\mu+i[A_\mu, A_\nu]$,
$A^{mn}_\nu=-i\langle \Psi_m|\partial_\mu |\Psi_n\rangle$.
Matrix $F_{\mu\nu}$ and $A_{\mu}$ are nonabelian Berry curvature and
nonabelian Berry connection for the ground state manifold, respectively.
$S^4$ is a closed 4D surface in the parameter space and
$\epsilon_{\mu\nu\rho\lambda}$ is Levi-Civita symbol. When
nodal lines or rings are observed, corresponding topological invariants are
computed.

Away from the origin of the parameter space, the single-particle ground state
becomes two-fold degenerate. Thus, for $N$ non-interacting bosons, there are
$N+1$ degenerate ground states and $C_2$ reads $N(N+1)(N+2)/6$~\cite{supnote}.  
Turning on interactions, results become completely different. 

{\it N Yang monopoles. } When $g>0$,
there are $N$ points on the $R_5$ axis, where the many-body ground state
becomes four-fold degenerate. At the origin, 
$\hat{K}=0$, there are four ways to distribute $N\in odd$ bosons in
four equivalent lattice sites to minimize the interaction energy, as shown in
Fig.~\ref{fig1}(b-c). Away from the origin, four-fold degenerate points also exist on the
$R_5$ axis.  All tunnelings in Eq.~(\ref{4site}) vanish on this axis, as
$R_{i\neq 5}=0$. 
Many-body eigenstates are simply Fock states.  The mismatch of onsite energies
$\epsilon_1-\epsilon_2=\epsilon_4-\epsilon_3$
could exactly compensate the penalty of interaction energy for moving one boson
from one lattice site to another. For example, for $N=3$ and $R_5=g$, states
$\ket{1,1,0,1}, \ket{1,0,1,1}, \ket{2,0,0,1},$ and $\ket{1,0,0,2}$ become
degenerate. For any $N$, the separation between
two nearest points is given by $\Delta R_5=g$.

Away from these four-fold degenerate points, the four Fock
states are no longer degenerate, and tunnelings become finite.
In the vicinity of each
degenerate point, we construct an effective model using the four nearly
degenerate states as the basis. Such effective model has exactly
the same formula as the single-particle Hamiltonian in Eq.~(\ref{4site})
except that $\epsilon_i$ and $t_{ij}$ are modified.
We show, e.g., the effective Hamiltonian near the
origin, in which the parameters read
\begin{align}
  \tilde{\epsilon}_i =(-1)^{\frac{N-1}{2}}\epsilon_i  \text{, } 
  \tilde{t}_{ij}=
  \begin{cases} t_{ij}(N+3)/4  \text{ for $N=$1, 5, \dots}\\
    t_{ij}^*(N+1)/4 \text{ for $N=$3, 7, \dots}
  \end{cases}
\end{align}
Thus, we conclude that each four-fold degenerate point
corresponds to a Yang monopole. 
A subtle difference between $N=1,5, ...$ and 
$N=3, 7, ...$ exists. As shown in Fig.~\ref{fig1}(b-c),
it is a particle and a hole that tunnels in
the effective Hamiltonian for these two cases, respectively. The
``charge" of the Yang monopole at the origin for $N=1,5, ...$ is $1$ and that
for $N=3, 7, ...$ is $-1$. Similarly, for a fixed $N$, with increasing 
distance to the origin, the ``charges" of monopoles 
alternate~\cite{supnote}. When all monopoles are enclosed, $C_2=1$.

{\it A giant Yang monopole.} For attractive interactions, only one monopole
exists in the parameter space,
and its ``charge" is $N^2$. At the origin, $|N,0,0,0\rangle$, $|0,N,0,0\rangle$,
$|0,0,N,0\rangle$, $|0,0,0,N\rangle$ are the four degenerate many-body ground
states, as all bosons prefer to occupy the same lattice site to minimize the
interaction energy. Away from the origin, an effective model, which has the same
formula as Eq.~(\ref{4site}), can be constructed. Since a single-particle
tunneling $t_{ij}$ moves one boson from one lattice site to another, it requires
$N$ steps of single-particle tunneling to couple these states. The
parameters in the effective Hamiltonian read
\begin{align}
  \tilde{\epsilon}'_i=N\epsilon_i\; \text{and}\;  
  \tilde{t}'_{ij}=c_Nt_{ij}^N/g^{N-1},
\end{align}
where $c_N$ is a function of $N$~\cite{supnote}.
Using this effective
model, we obtain that the ``charge" of the monopole is $N^2$
[Fig.~\ref{fig1}(d)]. The
superposition of the four Fock states actually forms a Schrodinger
cat state~\cite{cat0,catexp,cat1,cat2}. Though not stable for large $N$,
in a few-body system~\cite{jochim13,jochim132}, a small cat could exist in
laboratories such that a Yang monopole of ```charge" $N^2$ is observable.

$C_2$ for any closed surface is equal to the total ``charge'' of the monopoles
it encloses.
If a smooth deformation of the surface does not touch a Yang monopole,
$C_2$ remains unchanged.
We numerically calculate $C_2$ of 3 particles on a 4D sphere as a function
of the radius of the sphere.
Fig.~\ref{fig1}(e-f) show that $C_2$ is indeed given by the total
``charge'' of the monopole enclosed in the sphere. Note that $C_2$ 
is much smaller than that of non-interacting systems.
This is because an infinitesimal interaction reduces the $N+1$ fold degeneracy
of non-interacting systems to a two-fold one. 
Nevertheless, we have verified that, the total $C_2$ of the lowest $N+1$ bands
for weakly interacting systems is indeed the same as that for the corresponding
non-interacting systems.

{\it  3D topological defects. }
If the average particle number per site is an integer,
i.e., $N=4n$, where $n$ is a positive integer, the many-body ground state becomes
unique for $g>0$. 
This is best understood in the strongly
interacting regime. As bosons prefer to distribute evenly in the four lattice
sites to minimize the interaction energy, the unique ground state
cannot see any topological defects.  When $g<0$, the many-body ground
state is four-fold degenerate at the origin of the parameter space,  similar to
the case of odd particles. Away from the origin, an effective Hamiltonian can
be constructed in the same manner.  However, the resultant effective
Hamiltonian is distinct.
The effective coupling between the
Fock states, such as $|N,0,0,0\rangle$ and $|0, N,0,0\rangle$, now requires an
even number steps of single-particle tunnelings. In the single-particle
Hamiltonian 
in Eq.~(\ref{4site}), $t_{12}$ and $t_{34}$ have different
signs. 
This minus sign remains unchanged in the effective model for odd $N$, 
as both 
effective couplings, $\tilde{t}'_{12}$ and $\tilde{t}'_{34}$, are
proportional to odd powers of $t_{12}$ and $t_{34}$. 

For even particle numbers, the minus sign disappears. Completely different
topological defects arise. The effective Hamiltonian reads
\begin{equation}
  \hat{H}_\text{eff}=a \tilde{\tau}_z\otimes
  \tilde{\sigma}_z+b\tilde{\tau}_x\otimes I +c\tilde{\tau}_y \otimes I+d I
  \otimes \tilde{\sigma}_x +e I \otimes \tilde{\sigma}_y,
  \label{He}
\end{equation}
where 
$a=-2 R_5$, $b=-(R_1^2-R_2^2)/g$, $c=-2R_1R_2/g$,
$d=-(R_3^2-R_4^2)/g$, and $e=-2R_3R_4/g$ for $N=2$.
$\vec{\tilde{\sigma}}$ and $\vec{\tilde{\tau}}$ are two spin-1/2s, and $I$ the
identity matrix. The
eigenstates of $\tilde{\tau}_z\otimes \tilde{\sigma}_z$,
$|\uparrow\uparrow\rangle, |\uparrow\downarrow\rangle,
|\downarrow\uparrow\rangle, |\downarrow\downarrow\rangle$, correspond to
$|N,0,0,0\rangle$, $|0,N,0,0\rangle$, $|0,0,N,0\rangle$, $|0,0,0,N\rangle$. The
eigenenergy of $\hat{H}_\text{eff}$ reads 
\begin{equation}
  E=\pm\sqrt{a^2+
    (\sqrt{b^2+c^2}\pm \sqrt{d^2+e^2})^2},
    \label{en}
\end{equation}
which shows
that eigenstates become degenerate in certain 3D continuous manifolds.

\{{$\mathcal{M}_1$}: $R_1=R_2=0$\} and \{{\it $\mathcal{M}_1'$}:  $R_3=R_4=0$\},
both the ground and excited states are doubly degenerate. 

\{{$\mathcal{M}_2$}: $R_5=0$, $R_1^2+R_2^2=R_3^2+R_4^2$\}, the second and third
states are degenerate, and the ground state (the fourth state) is unique. 

As Kramers theorem does not apply to even number of spin-3/2s,
the even-fold degeneracy is not guaranteed and $\mathcal{M}_2$ is possible here.
These three manifolds intersect
at the origin of the 5D parameter space. Away from them, there is no degeneracy.
{\it $M_2$} signifies the vanishing gap between the lowest and the highest two
states on any closed 4D surface. Thus, $C_2$ is no longer  appropriate  
to characterize the topological defects.  Each manifold 
is characterized by its own corresponding topological invariant.
Meanwhile, the projections of them in lower
dimensions lead to knotted nodal lines and rings. 

Since {\it $\mathcal{M}_1$} and {\it $\mathcal{M}_2$} are 3D defects in a 5D
parameter space, a 1D loop can be defined without intersecting them. 
We calculate the Berry phase
$\gamma_m=-{i}\oint_{M} d{\bf R}\cdot \langle \Psi_m
  |\nabla_{\bf R} |\Psi_m\rangle$ 
for the $m$-th eigenstate $|\Psi_m\rangle$, where $M$ denotes a closed loop
in the parameter space.
For any loop that does not interlock the defects,
i.e., loop that can shrink to a single point without
closing the gap, $\gamma_m=0$. For a loop interlocking the defects, whether
$\gamma_1+\gamma_2=0$ or $\pi$ (or their multiples), defines a $Z_2$ index 
$\zeta_1$ for the
defects~\cite{liang15}. For {\it $M_1$} and {\it $M_1'$}, we find
that $\gamma=N\pi$ for all eigenstates and $\zeta_1=0$.
For {\it $M_2$}, we find that
 $\gamma=(0,\pi,\pi,0)$ for each eigenstate and $\zeta_1=1$.

To better visualize this $Z_2$ invariant, we project { $\mathcal{M}_1$},
{$\mathcal{M}_1'$} and {$\mathcal{M}_2$} to lower dimensions, i.e., reducing the
dimension by fixing the values of certain parameters. Defining $\vec{m}=(d,e)$
and $|\vec{m}|=\sqrt{d^2+e^2}=(R_3^2+R_4^2)/|g|$, the eigenenergies in
Eq.~(\ref{en}) read $E=\pm\sqrt{a^2+ (\sqrt{b^2+c^2}\pm |\vec{m}|)^2}$.
Interestingly,
this energy spectrum is identical to the one used to study nodal rings in
electronic systems~\cite{liang15}. 
As shown in Fig.~\ref{fig2},
\begin{figure} \centering
  \includegraphics[angle=0,width=0.45\textwidth]{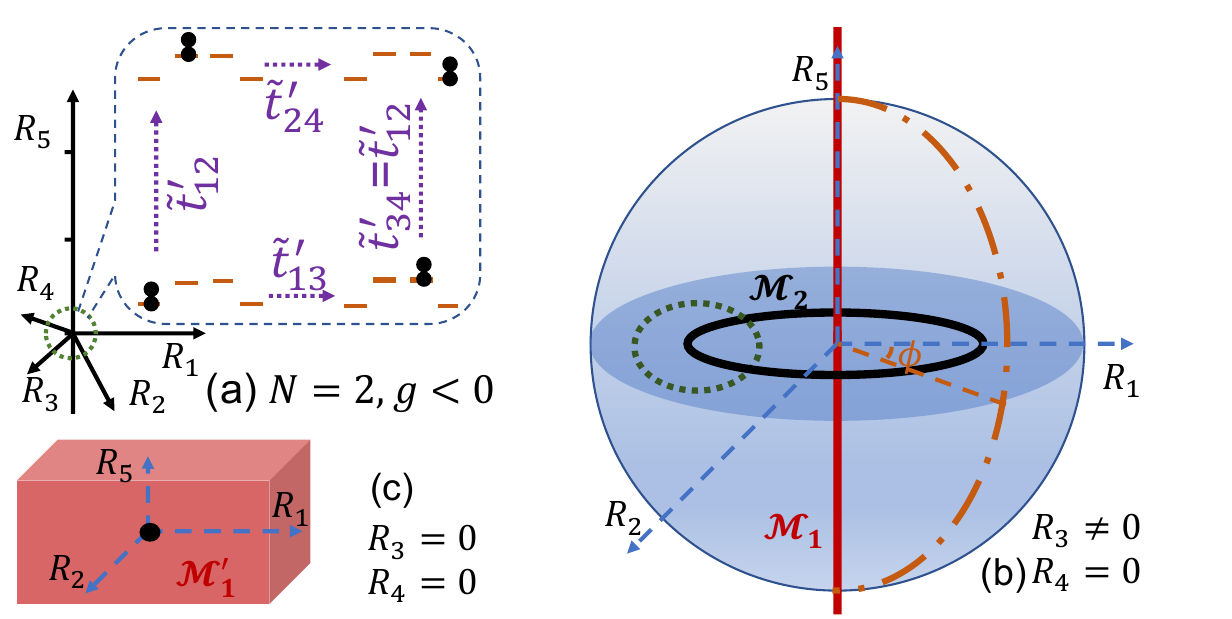}
  \caption{
    (a) The effective Hamiltonian for two particles with $g<0$. 
    (b-c) shows the projections of the defects in a 3D subspace 
    with fixed $R_3$ and $R_4$.  
    (b) The red line (black ring) shows the projection of
    $\mathcal{M}_1$ ($\mathcal{M}_2$). The green dotted circle
    (blue 2D sphere) is used to calculate $\zeta_1$ ($\zeta_2$). 
    The dash-dotted longitude connecting the north and south poles 
    defines a $\phi$-dependent Wilson line. 
    (c) When $\mathcal{M}_2$ reduces to a point at the origin,
    $\mathcal{M}_1'$ occupies the entire 3D subspace (red box).
  }\label{fig2}
\end{figure}
for any finite $|\vec{m}|$, $\mathcal{M}_1$ becomes
an infinite nodal line, and $\mathcal{M}_2$ becomes a nodal ring with radius
$|\vec{m}|$. $\mathcal{M}_1'$ does not show up in this
subspace. The dashed circles that 
interlock the nodal ring or line allow
one to compute $\gamma$. Decreasing $|\vec{m}|$, the nodal ring shrinks and
the nodal line remains unchanged. When $|\vec{m}|=0$, the nodal ring reduces
to a single point at the origin, and the gap does not open. In particular,
this whole 3D subspace precisely becomes $\mathcal{M}_1'$ and the
eigenenergies are 
two-fold degenerate everywhere. For this particular set of parameters,
$\hat{H}_\text{eff}$ describes a quantum spin Hall effect, as $\sigma_z=\pm 1$
corresponds to two opposite effective magnetic fields acting on $\vec{\tau}$.
When $\vec{m}$ changes sign, the nodal ring appears again. 

On any 2D sphere that does not touch {$\mathcal{M}_2$}, 
the lowest two eigenstates are separated from the rest.  
On such sphere, the projection to the lowest $k$ states, which are separated
from the higher $l$ states, establishes another topological invariant,
$\zeta_2$, of the nodal ring~\cite{hatcher02,liang15}. To be explicit,
Wilson lines connecting the north and south pole along a half longitude depend
on the polar angle $\phi$, as shown in Fig.(\ref{fig2}). Such $\phi$ dependence
allows one to define a winding number $n_w$.
For generic $k,l>2$, $\zeta_2=\mod(n_w, 2)$ defines a $Z_2$ index.
In our system, $k=2$, and the winding number becomes a $Z$
index~\cite{liang15,dai16}. 
For our effective model $\hat{H}_\text{eff}$, we find that $\zeta_2=1$
for any 2D sphere that encloses the nodal ring;
Otherwise, $\zeta_2=0$. Thus, the nodal ring defines a
topological phase transition where $\zeta_2$ changes its value. For
repulsive interactions, {\it $M_1$} and {\it $M_1'$} switch with
{\it $M_2$}~\cite{supnote}.
While the topological defects are derived in the strongly interacting regime,
we numerically verified that they
hold even for weakly interacting systems. 

{\it Realizations in few-body systems.}
With increasing $N$, the
energy splitting between eigenstates decreases.
Moreover, due to the small scattering length $a_s$ and the extended
orbital wavefunction in the  NIST experiment,
interaction strength $g$ is very weak. For instance,
for $a_s=5$nm, $N=10^5$, trapping frequency $\omega=2\pi*70$Hz, 
$g\approx0.04$Hz,
which is too weak to have significant effects.
The main experimental results are well explained by
non-interacting pictures. Thus, to better
resolve these topological defects and the associated topological invariants,
experimentalists could use few-body systems to reduce $N$ and increase $g$. 

A 2D optical superlattice is a promising platform to realize Hamiltonians
in Eq.~(\ref{int}) and Eq.~(\ref{He}) in the real
space. Such superlattice divides the system into many isolated plaquettes,
each of which contains four sites. Currently available experimental techniques
allow experimentalists to dress and detect each individual plaque.
Many interesting few-body phenomena have been
explored~\cite{bloch12,haper0,pan16}. 
Using laser-assisted tunneling and magnetic field gradient, both the amplitude
and phase of the tunnelings can be engineered~\cite{haper0,haper1}.
The onsite potential can be tuned by superposing an additional 
1D lattice tilted by 45$^\circ$.
The Hamiltonian in Eq.~(\ref{spinmodel}) can then be delivered.
Turning on interactions, the effective Hamiltonian in 
Eq.~(\ref{He}) could then be explored. 
For instance, the interaction strength
is around $100\text{Hz}$ for Rb in optical lattices with laser wavelength
of 767nm and depth of $8E_R$, where $E_R$ is the characteristic energy scale
defined by the wavelength. Increasing the lattice depth or the scattering
length, $g$ can be further enhanced. Using realistic experimental parameters,
we find that  the previously discussed topological defects can indeed be
resolved~\cite{supnote}. Experimentalists can also realize Eq.~(\ref{He})
directly in non-interacting systems by engineering the inter-site couplings.
A few other approaches, including mesoscopic traps, optical tweezers,
ion traps and superconducting circuits, can also be used to study few-body
physics related to our work~\cite{supnote,jochim13,jochim132}.

A unique advantage of ultracold atoms is that topological defects and the
associated topological invariants can be directly probed. To measure 
$\zeta_1$, the local Berry curvature could be measured to
extract the Berry phase accumulated in a 1D
loop~\cite{pnas12,berry12,zakexp13,berry14,berry15,kolodrubetz16,nistarXiv}.
To measure $\zeta_2$, Wilson lines can be measured in the
same manner in Ref.~\cite{wilsonexp16}.

In the NIST experiment on spinor BEC, inter-spin interactions exist.
Thus, we need to consider generic interactions
$\sum_ig_in_i^2+\sum_{i<j}g_{ij}n_in_j$, where $g_i$ ($g_{ij}$) is the intra
(inter) spin interaction strength. If such interactions preserve time reversal
symmetry, our main results remain unchanged. For interactions that break time
reversal symmetry, even richer physics regarding topological defects
arises~\cite{supnote}. 

In conclusion, we have shown that interactions give rise to emergent topological
defects distinct from those seen by each individual particle. Depending on the
total particle number and interaction strength, either giant Yang monopoles of
multiple charges or 3D continuous topological defects emerge.
Such topological defects can be accessed in current experiments, in particular,
those on few-body systems. While Dirac monopoles control many 2D and 3D
topological matters, Yang monopoles and $C_2$ are crucial for topological
quantum phenomena in high dimensions, including 4D quantum Hall effects.
Nodal lines and nodal rings as continuous topological defects also provide
physicists unprecedented topological quantum matters. 
We hope that our work will stimulate more studies on using ultracold atoms to 
create and measure topological defects in high dimensional interacting systems.

\begin{acknowledgments}
QZ acknowledges useful discussions with T.L. Ho, I. Spielman, S. Sugawa, and
C. Fang.
YY acknowledges useful discussions with C. Li.  This work is supported by
startup funds from Purdue University. 
\end{acknowledgments}


%

\pagebreak
\widetext
\begin{center}
\textbf{\large 
  Supplemental Material of ``Yang monopoles and emergent three-dimensional topological defects in interacting bosons"
}
\end{center}
\renewcommand{\theequation}{S\arabic{equation}}
\renewcommand{\thefigure}{S\arabic{figure}}
\renewcommand{\thetable}{S\arabic{table}}

\maketitle
\onecolumngrid
\section{definition of $C_2$}

The non-abelian Berry connection $A^{mn}_\nu$, non-abelian Berry curvature $F^{mn}_{\mu\nu}$, and non-abelian second Chern number are defined as~\cite{liarXiv}.
\begin{align}
A^{mn}_\nu&=-i\langle \Psi_m|\partial_\mu |\Psi_n\rangle, \,\,\,\,F^{mn}_{\mu\nu}=\partial_\mu
A^{mn}_\nu-\partial_\nu A^{mn}_\mu+i[A_\mu, A_\nu]^{mn}\\
  C_2&=\frac{1}{32\pi^2}\int_{S^4} d {\bf
  R}\epsilon_{\mu\nu\rho\lambda}(\text{Tr}
  [F_{\mu\nu}F_{\rho\lambda}]-\text{Tr}[F_{\mu\nu}]\text{Tr}[F_{\rho\lambda}]).
  \label{C2}
\end{align}
\section{non-interacting $C_2$}
 Single-particle ground state of the Hamiltonian in Eq.~(1) of the main text is doubly degenerate. Denote the two
states as $\psi_a$ and $\psi_b$, and define $A_{\mu\nu}^{mn}=-i\braket{\partial_{\mu}\psi_m|\partial_{\nu}\psi_n}$,
the Berry curvature reads
\begin{equation}
  F_{\mu\nu}^{mn}=A_{\mu\nu}^{mn}-A_{\nu\mu}^{mn}+i[A_{\mu},A_{\nu}].
  \label{}
\end{equation}

Because of the time-reversal symmetry, the traces of the Berry connection and the Berry curvature are zero,
\begin{align}
A_{\mu}^{aa}=-A_{\mu}^{bb},  \,\,\,\,\,\, \text{Tr} F_{\mu \nu}=0.
  \label{}
\end{align}
We also have $ A_{\mu\nu}^{aa}=A_{\nu\mu}^{bb}$.  Defining
\begin{align}
  \tilde{A}_{\mu\nu}=\left(\begin{array}{cc} A_{\mu\nu}^{aa} & A_{\mu\nu}^{ab}\\A_{\mu\nu}^{ba} & -A_{\mu\nu}^{aa}\end{array}\right)\\
    \tilde{F}_{\mu\nu}=\tilde{A}_{\mu\nu}+iA_{\mu}A_{\nu},
  \label{}
\end{align} and using the property of the Levi-Civita symbol,
we express the second Chern number using an alternative form,
\begin{equation}
  C_2^o(1)
  =\frac{1}{32\pi^2}\int_{S^4} d {\bf
  R}\epsilon_{\mu\nu\rho\lambda}\text{Tr}
  [F_{\mu\nu}F_{\rho\lambda}]
  =\frac{1}{8\pi^2}\int_{S^4} d {\bf
  R}\epsilon_{\mu\nu\rho\lambda}\text{Tr}
  [\tilde{F}_{\mu\nu}\tilde{F}_{\rho\lambda}].
  \label{C2}
\end{equation}
The above results can be generalized to $N$ particles.

For $N$ non-interacting bosons, there are $N+1$ degenerate ground states.
Using Fock states as basis states, the ground state reads $\ket{N-i,i}$, 
where $i$ takes the values of $0, 1, 2,\dots, N$; here
$\ket{N-i,i}$ represent the $N-i$ bosons in state $\psi_a$ and $i$ bosons in
state $\psi_b$.
The Berry connection $A_\mu$ and matrix $A_{\mu\nu}$ are tridiagonal in the Fock
state basis.
\begin{align}
 & A_{\mu}^{i-1,i}=-i\braket{N-i+1,i-1|\partial_\mu|N-i,i}=\sqrt{i(N-1+1)} A_{\mu}^{ab}\\
  &A_{\mu}^{i,i-1}=-i\braket{N-i,i|\partial_\mu|N-i+1,i-1}=\sqrt{i(N-1+1)} A_{\mu}^{ba}\\
  &A_{\mu}^{i,i}=-i\braket{N-i,i|\partial_\mu|N-i,i}=(N-i) A_{\mu}^{aa}+i A_{\mu}^{bb}=(N-2i)
  A_{\mu}^{aa}.
  \label{}
\end{align}
The same holds for the $\tilde{A}_{\mu\nu}$ matrix,
\begin{align}
&  \tilde{A}_{\mu\nu}^{i-1,i}= -i\braket{N-i+1,i-1|\overleftarrow{\partial_{\mu}}\overrightarrow{\partial_{\nu}}|N-i,i}=\sqrt{i(N-1+1)} A_{\mu\nu}^{ab}\\
&  \tilde{A}_{\mu\nu}^{i,i-1}=-i \braket{N-i,i|\overleftarrow{\partial_{\mu}}\overrightarrow{\partial_{\nu}}|N-i+1,i-1}=\sqrt{i(N-1+1)} A_{\mu\nu}^{ba}\\
 & \tilde{A}_{\mu\nu}^{i,i}= (N-2i)
  A_{\mu\nu}^{aa}.
  \label{}
\end{align}
Note that $A_{\mu}^{xy}$ and $A_{\mu\nu}^{xy}$ are for single particle if
$x,y$ is $a$ or $b$ and are for $N$ particles elsewise.
Same as that for $N=1$, the second Chern number for arbitrary $N$ can also be calculated using Eq.~(S5).

We break the integrand $\epsilon_{\mu\nu\rho\lambda}\text{Tr}
  [\tilde{F}_{\mu\nu}\tilde{F}_{\rho\lambda}]$ into two parts,
$\epsilon_{\mu\nu\rho\lambda}\text{Tr}
[\tilde{F}_{\mu\nu}\tilde{A}_{\rho\lambda}]$ and 
$2i\epsilon_{\mu\nu\rho\lambda}\text{Tr}
[\tilde{A}_{\mu\nu}A_{\rho}A_{\lambda}]$
(note that $\epsilon_{\mu\nu\rho\lambda}\text{Tr}
[A_{\mu}A_{\nu}A_{\rho}A_{\lambda}]$ is 0).
Writing down the trace explicitly and expressing the two terms in terms of the 
single particle quantities, we obtain
\begin{align}
\text{Tr}
[\tilde{A}_{\mu\nu}\tilde{A}_{\rho\lambda}]
&=\sum_i \tilde{A}_{\mu\nu}^{ii}\tilde{A}_{\rho\lambda}^{ii}
+\sum_i \tilde{A}_{\mu\nu}^{i,i+1}\tilde{A}_{\rho\lambda}^{i+1,i}
+\sum_i \tilde{A}_{\mu\nu}^{i,i-1}\tilde{A}_{\rho\lambda}^{i-1,i}\\
&=\sum_{i=0}^{N}(N-2i)^2 \tilde{A}_{\mu\nu}^{aa}\tilde{A}_{\rho\lambda}^{aa}+
\sum_{i=0}^{N-1}(i+1)(N-i) 
\left(\tilde{A}_{\mu\nu}^{ab}\tilde{A}_{\rho\lambda}^{ba}+
  \tilde{A}_{\mu\nu}^{ba}\tilde{A}_{\rho\lambda}^{ab}\right)
  \label{}
\end{align}
\begin{align}
\text{Tr}
[\tilde{A}_{\mu\nu}A_{\rho}A_{\lambda}]
=
&\sum_i \tilde{A}_{\mu\nu}^{ii}
\left(A_{\rho}^{i,i+1}A_{\lambda}^{i+1,i}
+A_{\rho}^{i,i-1}A_{\lambda}^{i-1,i}\right)\\
&+\sum_i \tilde{A}_{\mu\nu}^{i,i-1}
\left(A_{\rho}^{i-1,i-1}A_{\lambda}^{i-1,i}
+A_{\rho}^{i-1,i}A_{\lambda}^{i,i}\right)\\
&+\sum_i \tilde{A}_{\mu\nu}^{i,i+1}
\left(A_{\rho}^{i+1,i+1}A_{\lambda}^{i+1,i}
+A_{\rho}^{i+1,i}A_{\lambda}^{i,i}\right)\\
=&\sum_{i=0}^{N}(N-2i)(N-i)(i+1)\left(A_{\mu\nu}^{aa}A_{\rho}^{ab}A_{\lambda}^{ba}+A_{\mu\nu}^{ba}A_{\rho}^{aa}A_{\lambda}^{ab}+A_{\mu\nu}^{ab}A_{\rho}^{ba}A_{\lambda}^{aa}\right)\\
&+\sum_{i=0}^{N}(N-2i)(N-i+1)i\left(A_{\mu\nu}^{aa}A_{\rho}^{ba}A_{\lambda}^{ab}+A_{\mu\nu}^{ba}A_{\rho}^{ab}A_{\lambda}^{aa}+A_{\mu\nu}^{ab}A_{\rho}^{aa}A_{\lambda}^{ba}\right)
  \label{}
\end{align}
Using the identity 
\begin{align}
  \frac{1}{6}N(N+1)(N+2)
  =-\sum_{i=0}^{N}(N-2i)(N-i+1)i,
  \label{}
\end{align}
we write $C_2^0(N)$ using $C_2^0(1)$,
\begin{equation}
  C_2^0(N)=\frac{1}{6}N(N+1)(N+2)C_2^0(1)=\frac{1}{6}N(N+1)(N+2).
  \label{}
\end{equation}
  
\section{location and charge of 
      Yang monopoles for odd $N$ and positive $g$}
Along the $R_5$ axis, all off-diagonal couplings disappear and Fock
states become the eigenstates. Varying $R_5$, we find $N$ possible
points on this axis where there exist four-fold degeneracy.
\subsection{(a) $N=4m+1$ for integer $m$}
The four states, $\ket{m-l+1,m+l,m+l,m-l},  \ket{m-l,m+l+1,m+l,m-l},$  $\ket{m-l,m+l,m+l+1,m-l},$ and $\ket{m-l,m+l,m+l,m-l+1}$ are degenerate when $R_5=-2lg$, where $l=-m,-m+1,\dots,m-1,m$.
The effective Hamiltonian reads
\begin{align}
  \tilde{\epsilon}_i &=\epsilon_i\\
  \tilde{t}_{ij}&=t_{ij}\sqrt{(m-l+1)(m+l+1)}.
\end{align}
The extra Bose enhancement factor comparing to the single particle Hamiltonian
does not change the second Chern number, so $C_2=1$.

At $R_5=-(2l+1)g$, the other four states, $\ket{m-l-1,m+l+1,m+l+1,m-l},  \ket{m-l,m+l,m+l+1,m-l},$  $\ket{m-l,m+l+1,m+l,m-l},$ and $\ket{m-l-1,m+l+1,m+l+1,m-l}$ are degenerate, where $l=-m,-m+1,\dots,m-2,m-1$.
The effective Hamiltonian reads
\begin{align}
  \tilde{\epsilon}_i &=-\epsilon_i\\
  \tilde{t}_{ij}&=t_{ij}^*\sqrt{(m-l)(m+l+1)}.
\end{align}
The extra Bose enhancement factor together with an additional phase factor $e^{i\pi}$ comparing to the single particle Hamiltonian
does not change the second Chern number but flipping the sign of the diagonal terms changes $C_2$ to $-1$.
\subsection{(b) $N=4m+3$ for integer $m$}
At $R_5=-2lg$, the four states,  $\ket{m-l,m+l+1,m+l+1,m-l+1}$, $\ket{m-l+1,m+l,m+l+1,m-l+1}$, $\ket{m-1,m+l+1,m+l,m-l+1},$ and $\ket{m-l+1,m+l+1,m+l+1,m-l}$ are degenerate, where $l=-m,-m+1,\dots,m-1,m$.
The effective Hamiltonian reads
\begin{align}
  \tilde{\epsilon}_i &=-\epsilon_i\\
  \tilde{t}_{ij}&=t_{ij}^*\sqrt{(m-l+1)(m+l+1)}.
\end{align}
Similar to case (a), $C_2=-1$.

At $R_5=-(2l+1)g$, the four states,  $\ket{m-l+1,m+l+1,m+l+1,m-l},  \ket{m-l,m+l+2,m+l+1,m-l},$  $\ket{m-l,m+l+1,m+l+2,m-l},$ and $\ket{m-l,m+l+1,m+l+1,m-l+1}$ are degenerate, where $l=-m-1, -m, \dots, m-1,m$.
The effective Hamiltonian reads
\begin{align}
  \tilde{\epsilon}_i &=\epsilon_i\\
  \tilde{t}_{ij}&=t_{ij}\sqrt{(m-l+1)(m+l+2)}.
\end{align}
Similar to case (a), $C_2=1$.

\begin{figure} \centering
  \includegraphics[angle=0,width=0.78\textwidth]{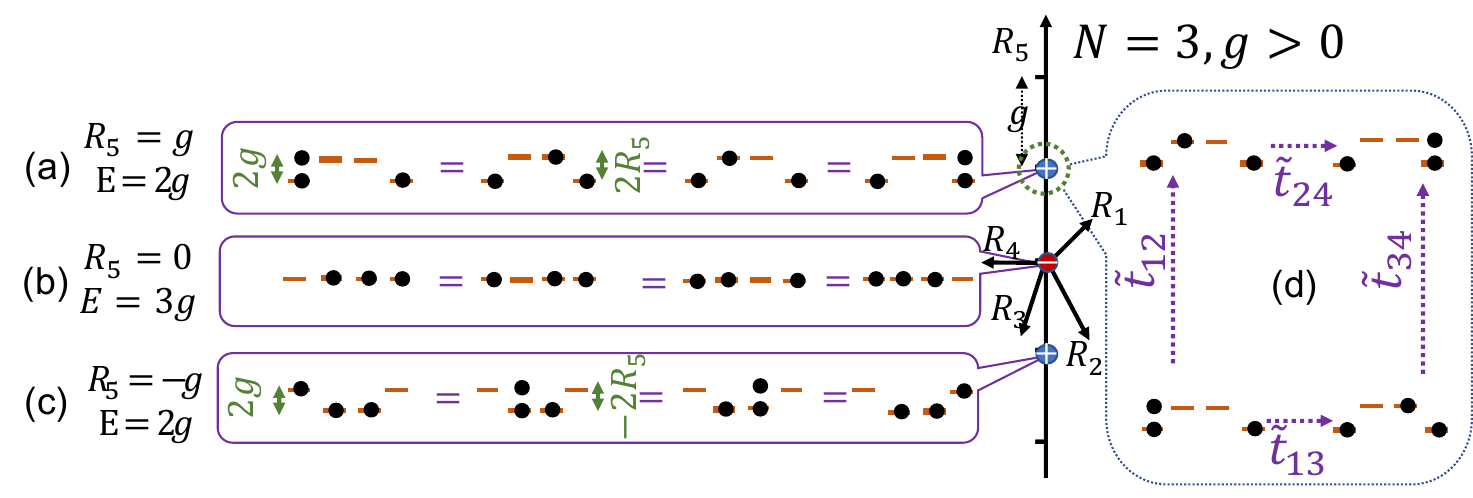}
  \caption{(Color online)
    Schematic of Yang monopoles for 3 particles with positive interaction
    strength $g$.
    Blue (red) spheres show the positively (negatively) charged monopoles. 
    Charges are denoted inside the sphere.
    The insects (a), (b), and (c) show the degenerate states at
    $R_5=g, 0,$, and $-g$,respectively; here
    $R_1=R_2=R_3=R_4=0$.
    The insect (d) shows the effective Hamiltonian in the vicinity of $R_5=g,
    R_1=R_2=R_3=R_4=0$.
    Orange solid lines and black dots represent single particle states and bosons, respectively.  
    Purple dotted arrows show effective couplings.
  }
  \label{figs1}
\end{figure}

For example, Fig.~\ref{figs1}(d)
illustrates the effective Hamiltonian for $N=3$ and $R_5=g$.
The onsite energy different balances off the repulsive energy for the sites
with two particles. Thus, all the four states has the same energy and as the
coupling $R_1,R_2,R_3,R_4$ approaches zero, the four states becomes the
eigenstates and four-fold degeneracy emerges.

\section{expression for $c_N$}
At the origin, $|N,0,0,0\rangle$, $|0,N,0,0\rangle$,
$|0,0,N,0\rangle$, $|0,0,0,N\rangle$ are the four degenerate many-body ground
states. Treating the $|R|$ as a small parameter, the lowest order coupling
reads
\begin{equation}
  \braket{N,0,0,0|\hat{H}_\text{eff}|0,N,0,0}=\frac{\prod_{i=0}^{N-1}\braket{N-i,i,0,0|\hat{K}|N-i-1,i+1,0,0}}{\prod_{i=1}^{N-1}\left(\braket{N-i,i,0,0|\hat{U}|N-i,i,0,0}-\braket{N,0,0,0|\hat{U}|N,0,0,0}\right)}.
  \label{<++>}
\end{equation}
Other terms can be written similarly. Comparing with the expression in the
main text, we have
\begin{equation}
  c_N=\frac{\prod_{i=0}^{N-1}\sqrt{(N-i)(i+1)}}{\prod_{i=1}^{N-1}\left[(N-i)^2+i^2-N^2\right]}.
  \label{}
\end{equation}

\section{topological invariants}

The Wilson line  $W_{ij}(\phi)$ is defined by $W_{ij}(\phi)=\langle
\Psi_{i}(\pi,\phi)|\hat{W}|\Psi_j(0,\phi)\rangle$, where $
|\Psi_{i=1,2}(0,\phi)\rangle$ and  $|\Psi_{i=1,2}(\pi, \phi)\rangle$ are the
lowest two eigenstate states at the north and south poles, respectively,
$\phi\in[0,2\pi)$ is the azimuthal angle, and
$\hat{W}=\prod_{k=1}^\mathcal{N}\hat{P}_k$,
$\hat{P}_k=\sum_{i=1}^2|\Psi_{i}(\theta_k,\phi)\rangle\langle
\Psi_{i}(\theta_k,\phi)|$ is the projection operator in the $k$th step if we
divide the longitude to $\mathcal{N}$ steps.  Physically, $W_{ij}(\phi)$
characterizes the probability of occupying either eigenstate at the south pole
if the initial state is an arbitrary superposition of the eigenstate at the
north pole after an adiabatic evolution along the longitude.
$W_{ij}(\phi)$ is an orthogonal matrix and can be made real after an
appropriate unitary transformation on $\hat{H}_\text{eff}$ (rotate $\hat{H}_\text{eff}$ to be real). 
Thus,  $\xi(\phi)=W_{11}(\pi/2, \phi)+iW_{12}(\pi/2, \phi)$ defines a winding
number 
 $n_w=-\frac{i}{2\pi}\int_0^{2\pi}
  \xi^*(\phi)\partial_{\phi}\xi(\phi)$.

\section{3D topological defects for even number of particles with repulsive interaction}
For later convenience, we
introduce $R_A$, $R_B$, $\phi_A$, and $\phi_B$, which are defined through
$R_A e^{-I \phi_A}=R_x- i R_y$ and 
$R_B e^{-I \phi_B}=R_z n_x- i R_z n_y$.
For $4m$ particles with strong repulsive interaction, the ground state is $\ket{m,m,m,m}$. No degeneracy is found.

For $4m+2$ particles with strong repulsive interaction, the ground state manifold has 6
states,
$\ket{m,m,m+1,m+1}, \ket{m,m+1,m,m+1}, \ket{m,m+1,m+1,m}, \ket{m+1,m,m,m+1}, \ket{m+1,m,m+1,m},$ and $\ket{m+1,m+1,m,m}$.
The Hamiltonian reads
\begin{equation}
\left(
\begin{array}{cccccc}
 0 & 0 & -e^{i \phi _B} (m+1) R_B & -e^{i \phi _B} (m+1) R_B & 0 & 0 \\
 0 & 0 & e^{i \phi _A} (m+1) R_A & -e^{i \phi _A} (m+1) R_A & 0 & 0 \\
 -e^{-i \phi _B} (m+1) R_B & e^{-i \phi _A} (m+1) R_A & 2 R_5 & 0 & -e^{i \phi _A} (m+1) R_A & -e^{i \phi _B} (m+1) R_B \\
 -e^{-i \phi _B} (m+1) R_B & -e^{-i \phi _A} (m+1) R_A & 0 & -2 R_5 & e^{i \phi _A} (m+1) R_A & -e^{i \phi _B} (m+1) R_B \\
 0 & 0 & -e^{-i \phi _A} (m+1) R_A & e^{-i \phi _A} (m+1) R_A & 0 & 0 \\
 0 & 0 & -e^{-i \phi _B} (m+1) R_B & -e^{-i \phi _B} (m+1) R_B & 0 & 0 \\
\end{array}
\right)
  \label{}
\end{equation}
Two of the states, i.e.
$(\ket{0,0,1,1}-e^{-2i\phi_B}\ket{1,1,0,0})/\sqrt{2}$ and $
  (\ket{0,1,0,1}+e^{-2i\phi_A}\ket{1,0,1,0})/\sqrt{2}$, have zero energy.
 Projecting out these two states, we write the effective Hamiltonian as a 4 by 4 matrix, 
  \begin{equation}
    \left(
\begin{array}{cccc}
 2 R_5 & \sqrt{2} e^{-i \phi _A} (m+1) R_A & -\sqrt{2} e^{-i \phi _B} (m+1) R_B & 0 \\
 \sqrt{2} e^{i \phi _A} (m+1) R_A & 0 & 0 & -\sqrt{2} e^{i \phi _A} (m+1) R_A \\
 -\sqrt{2} e^{i \phi _B} (m+1) R_B & 0 & 0 & -\sqrt{2} e^{i \phi _B} (m+1) R_B \\
 0 & -\sqrt{2} e^{-i \phi _A} (m+1) R_A & -\sqrt{2} e^{-i \phi _B} (m+1) R_B & -2 R_5 \\
\end{array}
\right).
    \label{}
  \end{equation}
  Rewriting the effective Hamiltonian using direct product of $\sigma$ and
  $\tau$ matrices, we obtain
  \begin{equation}
    R_5
    (\tau_z+\sigma_z)+\left(R_1'\sigma_x+R_2'\sigma_y-R_3'\tau_x+R_4'\tau_y+\tau_z(R_3'\tau_x+R_4'\tau_y)+\tau_z(R_1'\tau_x-R_2'\tau_y)\right),
    \label{<++>}
  \end{equation}
  where $R_1'=(m+1)(R_1-R_3)/\sqrt{2}$, $R_2'=(m+1)(R_2+R_4)/\sqrt{2}$ and
$R_3'=(m+1)(R_1+R_3)/\sqrt{2}$, and $R_4'=(m+1)(R_2-R_4)/\sqrt{2}$.
  Solving the effective Hamiltonian, the eigen energies reads
  \begin{equation}
    \pm\sqrt{2} (m+1) \sqrt{\pm\sqrt{\left(\left(R_A^2+R_B^2\right)+\frac{R_5^2}{(m+1)^2}\right){}^2-4  R_A^2 R_B^2}+R_A^2+R_B^2+\frac{R_5^2}{(m+1)^2}}.
    \label{}
  \end{equation}
Eigenenergies become degenerate in certain 3D continuous manifolds.

\{{$\mathcal{M}_1$}: $R_A=0$\} and \{{\it $\mathcal{M}_1'$}:  $R_B=0$\}, the second and the third
states are degenerate, and the ground state (the fourth state) is unique.

\{{$\mathcal{M}_2$}: $R_5=0$, $R_A=R_B$\}, both
the ground and excited states are doubly degenerate. 

For {\it $M_1$} and {\it $M_1'$}, we find
that the berry phase $\gamma=(0,2\pi,2\pi,0)$
 and $\zeta_1=0$ for the lowest two states.
For {\it $M_2$}, we find
that $\gamma=\pi$ for all eigenstates and $\zeta_1=0$ for the lowest two
states.

We also rewrite the manifolds in the main text using $R_A$ and $R_B$ and find that the two type of 3D manifolds are switched with each other as
the interaction strength changes from negative to positive. Furthermore, we have numerically verified that these degenerate manifolds extend to the
weakly interacting regime.

\section{linking number}
In the 3D subspace with finite $|\vec{m}|$, e.g., $R_3\neq0$ and $R_4=0$, $\mathcal{M}_1$ and $\mathcal{M}_2$ are knotted
nodal line and nodal ring.
Thus, a linking number can be defined as follows,
\begin{equation}
  L=\frac{1}{4\pi}\oint
  _{\mathcal{M}_1}\oint_{\mathcal{M}_2}\frac{{\bf r}_1-{\bf r}_2}{|{\bf
  r}_1-{\bf r}_2|^{\frac{3}{2}}}\cdot {d{\bf r}_1\times d{\bf r}_2}.
\end{equation}
A straightforward calculation shows that $L$ is always 1, which verifies that
the two nodal surfaces are knotted in the subspace.

\section{experimental schemes for exploring few-body physics}

There are a number of schemes to experimentally explore few-body physics related to the discussions in the main text, including optical superlattice, mesoscopic traps, optical tweezers, ion traps, and superconducting circuits.

\begin{figure} 
  \includegraphics[angle=0,width=0.78\textwidth]{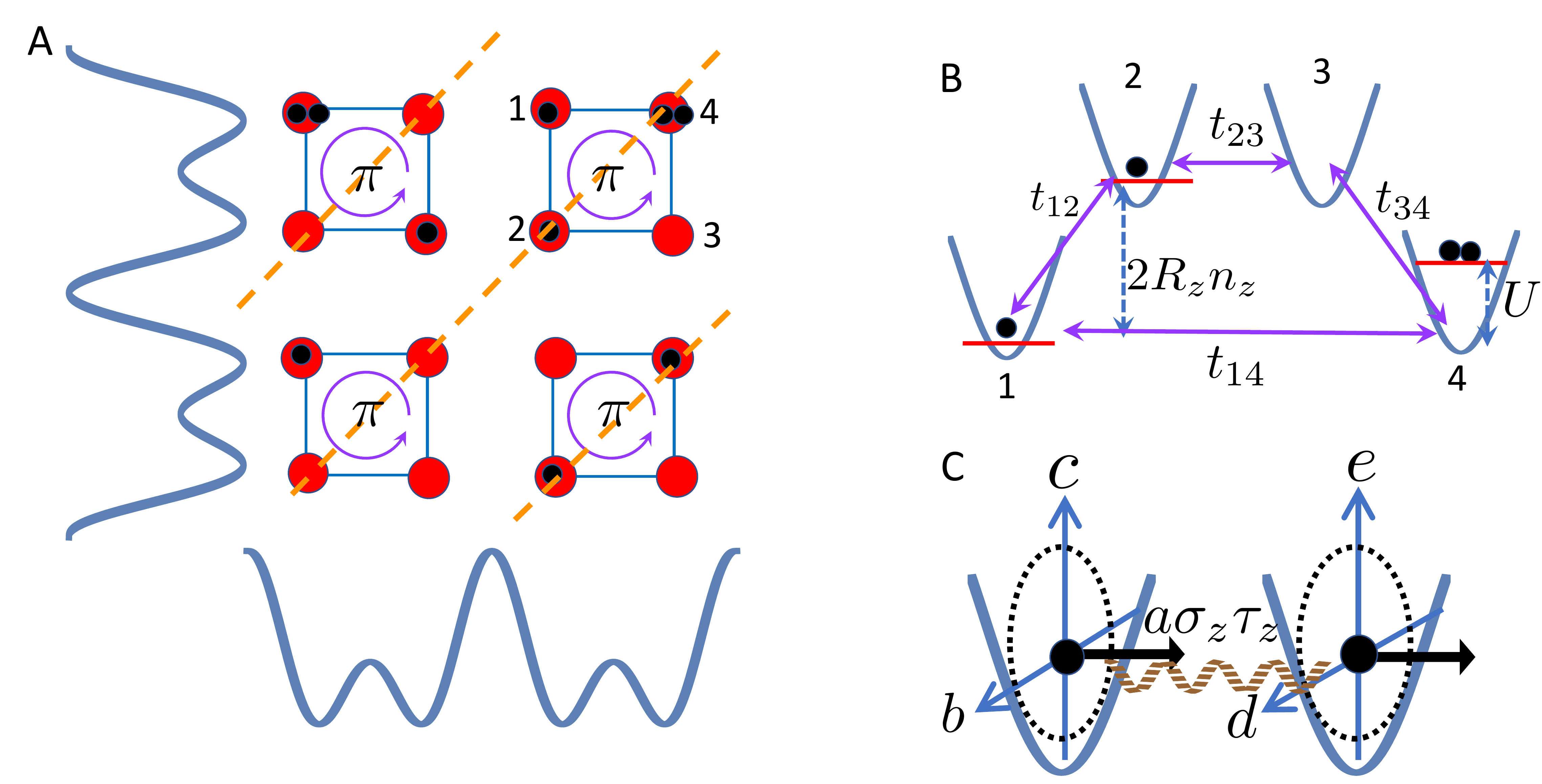}
  \caption{(Color online) (A) Red spheres represent lattice sites of an optical superlattice and blue curves represent lattice potentials. Black spheres represent atoms. Orange dashed lines represent an additional lattice potential that shifts the onsite energies of two sites in each plaquette. (B) A schematic of the tunnelings and onsite energy in four mesoscopic traps, optical tweezers, or each plaquette of an optical superlattice. (C) Our results also apply to two ion traps, each of which hosts a spin-$1/2$ (black arrows). }
  \label{figs2}
\end{figure}

As shown in Fig.~\ref{figs2}(A), a 2D optical superlattice is formed by a short and long lattice with wavelengths $\lambda_L=2\lambda_S$~\cite{haper0,haper1}.  Such superlattice divides the system into many plaquettes. When the energy barrier between different plaquettes is large enough, plaquettes are isolated from each other. Each contains a few particles, and can be dressed and probed individually. In particular, high resolution in-situ images allow experimentalists to measure precisely the particle number per plaquette. Optical superlattices have allowed physicists to explore many interesting few-body phenomena in two sites or four sites. 

To induce complex tunnelings among the four sites, either Raman dressing or shaking can be used. In the former approach, a field gradient can be applied in the diagonal direction to quench the bare tunnelings along both $x$ and $y$ directions~\cite{haper0,haper1}. Then a pair of Raman laser induces a complex laser assisted tunneling. In fact, it is not necessary to individually control the phase of each tunneling. The key requirement is that the total effective magnetic flux per plaquette is $\pi$, i.e., a particle accumulates a $\pi$ phase after finishing a closed loop composed of all four sites. Another approach is shaking the lattice~\cite{haldane,parker2013direct}. Theoretically, this is the same as the Raman dressed lattice. In practice, the advantage is that, no extra lasers are required.  The onsite energies can be tuned by the two-photon detuning or the shaking frequency. An additional 1D optical lattice aligned in the diagonal direction can provide extra controls of the onsite energy. Using typical experimental parameters for Rb with scattering length $\approx 5\text{nm}$ and laser wavelength $767$nm, we find that, interaction strength $g$ in such lattice can reach $100\text{Hz}$.  For other atoms, Feschbach resonance could further enhance $g$. 
$g$ is the characteristic energy scale of topological defects discussed here, for instance, the separation between different Yang monopoles in the parameter space and the energy gap in the effective models. With such $g$, topological defects can be easily resolved in current experiments. 

One could also use four mesoscopic traps~\cite{jochim13,jochim132}, or optical tweezers to realize a four-site system that is equivalent to a single plaquette, as shown in Fig.~\ref{figs2}(B). Two optical tweezers have been recently used to produce entangled pairs of atoms~\cite{kaufman2015entangling}. For instance, bringing four tweezers together and engineering the tunnelings between tweezers using external lasers,  a single four-site model could be realized.  Alternatively, one could stick to a single tweezer, and use four internal states of atoms. Theoretically, this would be equivalent to the NIST experiment. The advantage here is that, due to the strong confinement in an optical tweezer, the interactions could be much stronger. For instance, the typical confining potential of a single optical tweezer is $200\text{kHz}$. For $Rb$ with scattering length $a_s=5nm$ and $Cs$ with scattering length $a_s=91nm $, such confinement corresponds to an interaction strength $1\text{kHz}$ and $24\text{kHz}$, respectively. Thus, it will be much easier to explore interaction induced topological defects in optical tweezers.

Other quantum systems other than cold atoms can also be used to explore topological defects discussed in the main text. For instance, two nearby ion trap, each of which hosts a spin-$1/2$, can be used to realize the model in Eq.(7) of the main text, as shown in Fig. (S2.C). Here, $(b, c)$ and $(d,e)$ represent the magnetic field acting on the first and the second spin-$1/2$, which is denoted by $\vec{\sigma}$ and $\vec{\tau}$, respectively, in the $x-y$ plane. $a \sigma_z\tau_z$ is simply the Ising interaction. Similarly, two superconducting circuits may be used, as each circuit can be viewed as a spin-$1/2$~\cite{roushan2014observation,c114}. Therefore, our theoretical results can also be generalized to superconducting circuits.  

\section{generic interaction}

A generic interaction $\sum_ig_{i}n_i^2+\sum_{i\neq j}g_{ij}n_in_j$ leads to corrections to the effective Hamiltonian discussed in the main text.
As interactions transform to the onsite energy in the effective Hamiltonians constructed by four Fock states in Eq.~(1) and Eq.~(7) of the main text, nonuniform $g_i$ and nonlocal $g_{i\neq j}$ (or equivalently, the inter-spin interaction in the spin model) only lead to corrections in the diagonal terms. Most generically, the corrections can be written as $\hat{H}'_{eff}=\sum_i \delta_i H'_i$, where  
\begin{equation}
\begin{split}
&H_1'=\sum_{i=1}^4 \hat{a}_i^\dagger\hat{a}_i,\\
&H_2'=\hat{a}_1^\dagger\hat{a}_1+\hat{a}_2^\dagger\hat{a}_2-\hat{a}_3^\dagger\hat{a}_3-\hat{a}_4^\dagger\hat{a}_4,\\
&H_3'=\hat{a}_1^\dagger\hat{a}_1-\hat{a}_2^\dagger\hat{a}_2-\hat{a}_3^\dagger\hat{a}_3+\hat{a}_4^\dagger\hat{a}_4, \\
&H_4'=\hat{a}_1^\dagger\hat{a}_1-\hat{a}_2^\dagger\hat{a}_2+\hat{a}_3^\dagger\hat{a}_3-\hat{a}_4^\dagger\hat{a}_4.
\end{split}
\end{equation}
and $\delta_i$ corresponds to the strength of each type of perturbation.

$\delta_i$ depend on $g_i$ and $g_{ij}$. For instance,
the effective Hamiltonian near the origin of the parameter space for odd
number of particles reads 
 \begin{equation}
\left(
\begin{array}{cccc}
 -c_1 R_5+\delta _1+\delta _2+\delta _3+\delta _4 & -c_2 \left(R_1-i R_2\right) & -c_2 \left(R_3-i R_4\right) & 0 \\
 -c_2 \left(R_1-i R_2\right) & c_1 R_5+\delta _1+\delta _2-\delta _3-\delta _4 & 0 & -c_2 \left(R_3-i R_4\right) \\
 -c_2 \left(R_3-i R_4\right) & 0 & c_1 R_5+\delta _1-\delta _2-\delta _3+\delta _4 & c_2 \left(R_1-i R_2\right) \\
 0 & -c_2 \left(R_3-i R_4\right) & c_2 \left(R_1-i R_2\right) & -c_1 R_5+\delta _1-\delta _2+\delta _3-\delta _4 \\
\end{array}
\right)
\end{equation}
where $c_1=\tilde{\epsilon}_i/\epsilon_i$ and
$c_2=\tilde{t}_{ij}/t_{ij}$ [rewrite of Eq. (5) in the main text].
For three particles, $c_1=-1$, $c_2=1$, and 
\begin{align}
 \delta _1&= \frac{1}{4} \left(3 g_1+3 g_2+3 g_3+3 g_4+2 g_{12}+2 g_{13}+2 g_{14}+2 g_{23}+2 g_{24}+2 g_{34}\right)\\
 \delta _2&= \frac{1}{4} \left(-g_1-g_2+g_3+g_4-2 g_{12}+2 g_{34}\right)\\
 \delta _3&= \frac{1}{4} \left(-g_1+g_2+g_3-g_4-2 g_{14}+2 g_{23}\right)\\
 \delta _4&= \frac{1}{4} \left(-g_1+g_2-g_3+g_4-2 g_{13}+2 g_{24}\right).
  \label{<++>}
\end{align}
For systems with equal intraspin interaction and no interspin interaction, i.e., $g_i=g$, and $g_{ij}=0$, we obtain
$\delta_1=3g$ and $\delta_2=\delta_3=\delta_4=0$.
The Hamiltonian reduces to the unperturbed one. 

For generic cases where $\delta_i\neq 0$, these four corrections $H_i'$ can be classified into two categories.

{Category 1}: { $H_1'$ and $H_3'$ respect time reversal symmetry of the effective Hamiltonian. $H_1'$ only}
shifts the entire energy spectrum and does not change the
topological { defects. $H_3'$ only shifts the position of the topological defects as a finite $\delta_3$ simply changes the value of $R_5$ to $R_5$-$ \delta_3$.  Therefore, the shape of topological defects and other results in the main text remain unchanged.
 
{Category 2}: { $H_2'$ and $H_4'$ break the time reversal symmetry and new topological defects arise.}

When only $H_2'$ exists, i.e., $\delta_2\neq 0, \delta_4=0$, the energy spectrum reads
\begin{equation}
E_{eff}=\pm \sqrt{\pm2 \sqrt{\delta_2^2 \left(R_1^2+R_2^2+R_5^2\right)}+R_1^2+R_2^2+R_3^2+R_4^2+\delta_2^2+R_5^2}.
  \label{<++>}
\end{equation}
The two degenerate manifolds are obtained as follows.

1.${\mathcal{M}_1}$, the first (third) and the second (fourth) states become degenerate when $R_5=R_1=R_2=0$.

2.${\mathcal{M}_2}$, the second and the third states become degenerate when $R_5^2+R_1^2+R_2^2=\delta_2^2$ and $R_3=R_4=0$.

Apparently, ${\mathcal{M}_1}$ is an infinite 2D topological defect. In contrast,
${\mathcal{M}_2}$ becomes a finite 2D sphere, unlike ${\mathcal{M}_2}$ discussed in the main text, which extends to infinity. Thus, a finite $\delta_2$ breaks the four-fold degeneracy at the original Yang monopole but retains the degeneracy between the second and the third states on a 2D sphere, ${\mathcal{M}_2}$. 

Likewise, when only $H_4'$ exists, i.e., $\delta_4\neq 0, \delta_2=0$, we also obtain two degenerate manifolds,
 
1.${\mathcal{M}_1}$, the first (third) and the second (fourth) state become degenerate when $R_5=R_3=R_4=0$.

2.${\mathcal{M}_2}$, the second and the third states become degenerate when $R_5^2+R_3^2+R_4^2=\delta_4^2$ and $R_1=R_2=0$.

Again, ${\mathcal{M}_1}$ is an infinite 2D topological defect, and ${\mathcal{M}_2}$ is a finite 2D sphere.

When both $H_2'$ and $H_4'$ exit, i.e., $\delta_2\neq 0$ and $\delta_4\neq 0$, ${\mathcal{M}_2}$ can be written as, 
\begin{eqnarray}
\frac{R_5^2}{\delta_4^2}+\frac{R_3^2+R_4^2}{(\delta_4^2-\delta_2^2) } =1,\,\,\,\,\,\, |\delta_2|<|\delta_4| \label{el1}\\
\frac{R_5^2}{\delta_2^2 }+\frac{R_1^2+R_2^2}{(\delta_2^2-\delta_4^2) } =1,\,\,\,\,\,\, |\delta_2|>|\delta_4| \label{el2}
\end{eqnarray}
For generic $\delta_2\neq 0$ and $\delta_4\neq 0$, Eqs.~(\ref{el1}) and (\ref{el2}) describe an ellipsoid.
It is clear that, when $\delta_2=0$ or $\delta_4=0$, Eqs.~(\ref{el1}) and (\ref{el2}) reduce to the previous results.
When $|\delta_2|=|\delta_4|$, ${\mathcal{M}_2}$ becomes a line segment connecting $(0,0,0,0,-\delta_2)$ and $(0,0,0,0,\delta_2)$,
which signifies the transition from one ellipsoid in Eq.(\ref{el1}) to the other in Eq.(\ref{el2}).

In any case, if a 4D sphere encloses $\mathcal{M}_2$, $C_2$ remains unchanged. The reason is that, the second Chern number, as a topological invariant, is stable against small perturbations. Unless the perturbation is strong enough to close the energy gap between the second and the third energy eigenstates, $C_2$ remains the same. Thus, when the amplitudes of perturbations are much smaller than parameters of the single-particle Hamiltonian, for instance, the distance to the origin of the five-dimensional parameter space, $R$, results of $C_2$ in the main text remain unchanged.

We have also generalized the above results to the effective Hamiltonian for 3D continuous topological defects, i.e., the effective Hamiltonian in Eq.(7) of the main text. For two particles, if either  $\delta_2$ and $\delta_4$ is zero, $\mathcal{M}_1$ is defined as $R_1=R_2=R_5=0$. If both $\delta_2$ and $\delta_4$ are finite, $\mathcal{M}_1$ does not exist. In contrast, 
$\mathcal{M}_2$ always exists.  For finite $\delta_2$ and $\delta_4$,  $\mathcal{M}_2$ is shifted to $R_5=0$ and $R_1^2+R_2^2+\delta_4^2=R_3^2+R_4^2+\delta_2^2$.
Thus, small nonuniform $g_i$ and nonlocal interactions $g_{i\neq j}$ only lead to perturbative changes to the topological defects. 

\end{document}